# On the Formal Model for IEC 61499 Composite Function Blocks

Won-je KIM[1], Song-il CHA[2], Kyong-Jin Sok[3,4,*]

[1]Institute of Information Science & Technology, National Academy of Science, Pyongyang, 950003, DPR Korea

[2]Department of Computer Science, University of Sciences, Pyongyang, 950003, DPR Korea

[3]College of Mechanical and Electrical Engineering, Harbin Engineering University, Harbin, 150001, PR China.

[4]Institute of Information Technology, University of Sciences, Pyongyang, 950003, DPR Korea.

**Abstract**-The applications for IEC 61499 that is standard architecture for developing the applications of distributed control and measurement in factory automation, have the connected structure of the graphical elements called BFB(basic function block), SIFB(service interface function block) and CFB(composite function block). The research on the composite function block has been regarded as important issues in implementing hierarchy, multi-functionality and simplicity of software. Nowadays many researchers have been investigated IEC61499 in the fields of the software modeling composed of basic function block and service interface function block, the transformation from IEC61131 to IEC61499 and syntactic extension of ECC of basic function block. However, work related to the mathematical modeling for IEC61499 composite function block using in designing software with hierarchical structure is still lacking. This paper presents the mathematical model for the structure and execution analysis of IEC 61499 composite function blocks by using notation of the set theory. Also a subaplication configuration algorithm is suggested for the subapplication corresponding to the composite function block. Then its effectiveness through the computation experiment of several distributed control applications is shown. The proposed model can be used effectively as a basis for analyzing a runtime environment of a software tool for designing and developing the applications.

**Keywords**-IEC 61499, distributed control, modeling, composite function block, function block network, software development tool

I. Introduction

IEC 61499 is international standard which emerged at 2005 year and acknowledged officially and begins to apply for designing and building distributed control system [1~3]. For designing distributed control systems, this standard can be used to unify design and implementation of overall control system from field level to management. It also shows intuitively the configuration and behavior of hardware and software using function blocks. There are described rules that are necessary for designing and implementing the distributed control system, including architectures and function blocks, software tools.

Nowadays the IEC 61499 standard has been founded widely application in many areas of factory automation. In N. Kashyap et al. [4], presents the analysis and simulation results of the performance of fault location and isolation (FLI) using the IEC 61499, distribution automation standard in an automated power distribution feeder. In G. Zhabelova et al. [5], proposes the hybrid agent architecture specific to the power system automation domain using the industrial standards IEC 61850 and IEC 61499. In P. Lindgren et al. [6], proposes the method to provide safe end-to-end response times for distributed IEC 61499 applications communicating over switched Ethernet networks. In L. I. Pinto et al. [7], presents the results implementing the traffic light systems and PID controllers using IEC 61499 in ICARU-FB environment. In F. Andren et al. [8], a standard-based control approach for distributed energy resources is introduced and implemented. In Tao Penga et al [9], the function block technique, i.e. IEC 61499, is used for the development of energy demand models as it brings advantages such as modularity, encapsulation, extensibility and reusability.

The IEC 61499 standard defines execution processing of the each function block and simple scheduling function on network of the function blocks that compose of basic and service interface function blocks. Hence, the different implementations of the standard have made different assumptions how to execute the applications. As a result, the same application might behave differently when executed on different platforms [10]. In S. Panjaitan et al. [11], proposes the method for modeling of IEC 61499 functionality design using UML diagram. In M. Fletcher et al. [12], describes how function blocks can be used to build the holonic manufacturing systems. In Kim et al. [13], proposes the structural model of IEC 61499 application with composite functional blocks.

In [14, 15], is presented new design framework with hierarchical and concurrent novel extension of ECC (Execution Control Chart) for basic function blocks. In W. Dai et al. [16], proposes a new methodology of migration from IEC 61131-3 to IEC 61499 function blocks. In V. Dubinin et al. [17], a formal definition of an IEC 61499 application using the set theory notation is presented along with a semantic function block model that is based on a state-transition approach. This model is done by choosing the next transition from the enabled function block transitions according to the execution semantics modeled. Only sequential hypothesis execution semantic is mentioned in particular. In G. Čengić et al. [18], presents formal definitions of the application model using definitions of the types and instances of function blocks, application state space, external input and output sets. Only semantics of basic and service interface function blocks are of importance. In G. Čengić et al. [19], presents formal definitions on the three different execution models, based on formal model described in [18], buffered sequential execution model (BSEM), non-preempted multithreaded resource (NPTMR), cyclic buffered sequential model (CBEM), and shows its comparison results. In J. Carlson et al. [20], presents simple model of the applications with different function blocks, and analyzes its runtime behavior. Other attempts at formal modeling and verification of control related languages have been published [21], [22]. In D. L. Her et al. [21], sequential function charts (part of IEC 61131) are modeled and verified using time automata while in J.-R. Beauvais et al. [22] State charts are formally modeled using the SIGNAL language.

The previous researches of the application model are described mainly on the formal model for IEC 61499 applications that compose of basic and service interface function blocks, and are not discussed profoundly for the formal definitions and execution analysis on composite function blocks that can represent simply and hierarchically the application with complex structure and behavior.

This paper presents the formal definitions on the composite functional blocks of the IEC 61499 standard. The mathematical definitions have been used as a basis for implementation of a software tool for a runtime environment and formal verification.

This paper is organized as follows. Section II describes the structure of the function blocks in the IEC 61499 standard. Section III presents formal definitions on the composite function blocks of IEC 61499 standard. Section IV shows the result of computational experiments for IEC 61499 applications using proposed models. Finally, Section VI concludes this work.

II. Structure of function blocks in IEC 61499

The software architecture defined by the IEC 61499 standard is based on functional software units called function blocks that includes own algorithms and internal variables. There are three types of function blocks: basic function blocks, composite function blocks, and service interface function blocks. A basic function block executes an elemental control function, such as reading a sensor or setting the state of an actuator. The different function blocks may be combined together to encapsulate a higher-level control function, and such a combination is called a composite function block. The service interface function block provides the communication services among devices.
Figure 1 shows a structure of the components of a basic function block. The model distinguishes between events, data, and algorithms. The upper quadrant of Figure 1 shows the event stream, which executed the function block code. The arrival of an event executes one or more function block algorithms. After the execution of the algorithm, the function block will enable an output event. Output events are sent to other function blocks in the application and thus input events of relevant function blocks are occurred.
The data flow is shown in the lower quadrant of the function block in figure 1. Data inputs are provided either from physical devices or from other function blocks. At the time that an input event occurs, the relevant data input values are read and the appropriate algorithm is executed using the current data values. This results in a calculation or validation that yields data outputs of a function block. When the output event is triggered, the relevant data output is made available to another function block that continues the execution of the application.
Type name of the basic function block is an identifier that it can be used to uniquely identify the basic function block. The Execution Control Chart (ECC) of a basic function block determines which algorithm to execute based on the current input event and values of input, output and internal data variables. The example ECC in Fig. 2 states that if it is in initial state, STATE0, and input event EI is received, the ECC transfers to state STATE1 and schedules the algorithm named Alg for execution. After Alg has terminated, the output event EO is emerged, and the ECC transmits immediately to state STATE0 since the transition condition is "1" (true) and finishes execution.

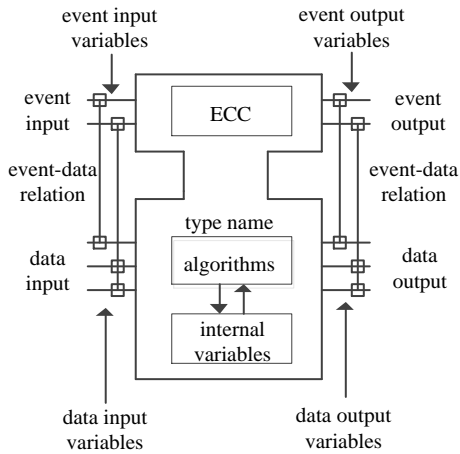
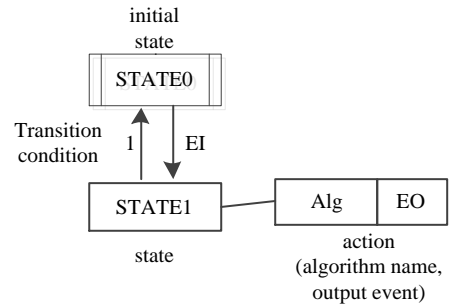

Figure 1. Components of a basic function block

Figure 2. Example of an execution control chart (ECC)

Another types of function blocks - a composite function block has its own interface(the input and output elements for events and data) as well as an basic function block and composes of a set of several basic and service interface, composite function blocks in which function blocks are connected by events and data. A composite function block that its elements of interface, i.e. input and output elements of events and data is connected with inner function blocks, can be used to perform higher level control function by making subapplication with inner function blocks.

A service interface function block is used to provide an application with access to services provided by an IEC 61499 execution resource, such as access to a communication network and the process under control.

We have briefly introduced about the IEC 61499 standard so far. The IEC 61499 applications can be executed using an IEC 61499 compliant runtime environment. The runtime environment may implement an IEC 61499 device composing of IEC 61499 execution resources. An important part of an execution resource is the scheduler. A scheduler's task is to provide each function block with an opportunity to execute when the block receives an event, i.e., a scheduler decides the block execution order. The following section presents the formal definitions on the composite function block of an IEC 61499 standard.

III. The formal definition on the composite function block

This section presents how a formal model on composite function block followed an IEC 61499 standard, is defined. In [18], it was presented the formal definitions for IEC 61499 applications that compose of basic and service interface function blocks, but not discussed on composite function blocks with hierarchical structure and behavior. This section presents formal definitions on the interface, type and instance of composite function block, and on subapplication and its execution corresponding the composite function block. Also, the composite function block that calculates function, $f(x, y) = x^2 - y^2$, is used as an illustrative example. Composite function block composes of the three basic function blocks that do the addition, subtraction and multiplication operation, respectively. The structure of composite function blocks, and of subapplication corresponding to it show in Figure 3 and Figure 4.

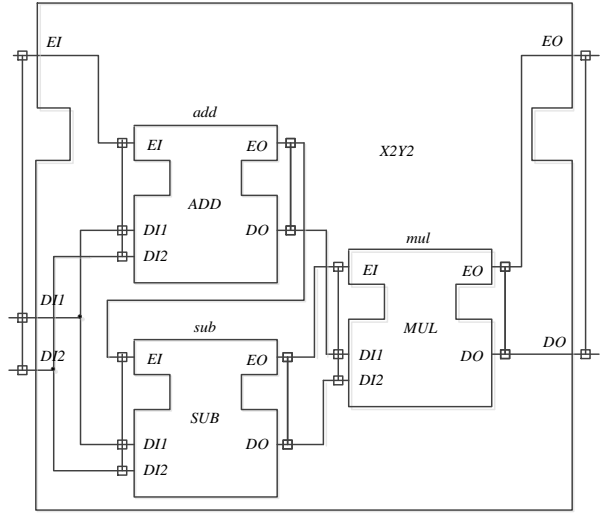

Figure 3. Structure of the composite function block $X2Y2$

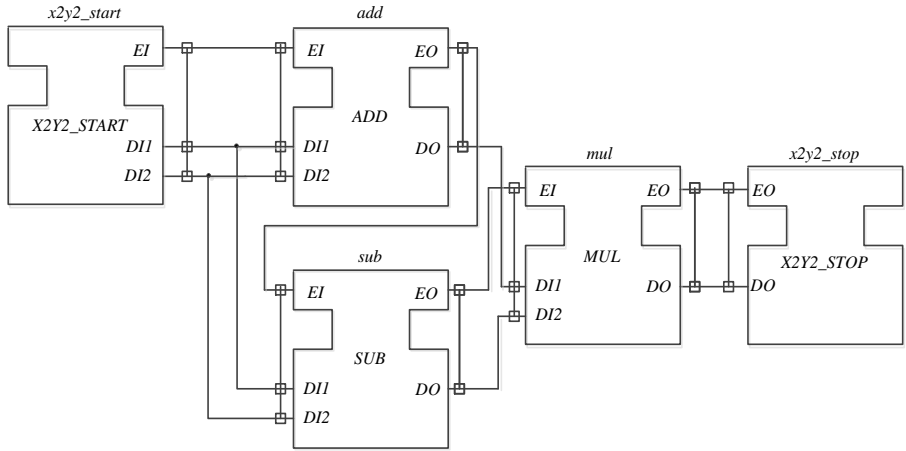

Figure 4. Structure of subapplication corresponding to composite function block

1) Structure of composite function block

First we define the interface common to all function blocks.

Definition 1(Function block interface): A function block interface is a 6-tuple defined as

$$g = \langle E_i, D_i, \omega_i, E_o, D_o, \omega_o \rangle$$

where $E_i, D_i, E_o,$ and $D_o$ are finite sets of input events, input data, output events and outputs data, respectively. $\omega_i \subseteq E_i \times D_i$ and $\omega_o \subseteq E_o \times D_o$ are the sets of input and output associations.

In composite function block of Figure 3, Interface $g$ is represented as follows.

$$g = \langle \{EI\}, \{EO\}, \tag{1}$$
$$\{\langle DI1,*\rangle, \langle DI2,*\rangle\}, \{\langle DO,*\rangle\}, \tag{2}$$
$$\{EI \to \langle DI1,*\rangle, \langle DI2,*\rangle\}, \{EO \to \langle DO,*\rangle\}\rangle \tag{3}$$

Line (1) describes the sets of input and output events of the block and line (2)- all values of input and output data, line (3) – the sets of input and output associations. Where symbol * indicates range of value of data input/output variables (for example, range of value of integer type variable).

Definition 2(Composite Function Block Type)

A composite function block type, $cfbt$, is a 3-tuple defined as

$$cfbt = \langle g, T, I \rangle$$

where $g$ is a composite function block interface, $T$ is a set of types of function blocks which a composite function block composes of them, $I$ is a set of instances of function blocks which a composite function block composes of them.

A set of inner function block types and instances that are components of composite function block type, $X2Y2$, in Figure 3, is represented as follows, respectively.

$$T = \{ADD, SUB, MUL\} \quad (4)$$
$$I = \{add, sub, mul\} \quad (5)$$

Composite function block instance is defined next.

Definition 3(Composite Function Block Instance)

Given composite function block type, $cfbt = \langle g, T, I \rangle$, with composite function block interface, $g = \langle E_i, D_i, \omega_i, E_o, D_o, \omega_o \rangle$, instance, $cfbi$, of composite function block type, $cfbt$, is a 5-tuple defined as

$$cfbi = \langle n, c_e, c_d, d_i, d_o \rangle$$

where $n$ is instance name, $c_e$ is a set of output event connections, $c_d$ is a set of input data connections, $d_i$ is an initial value set of input data variables $(d_i \in 2^{D_i})$, $d_o$ is an initial value set of data output variables $(d_o \in 2^{D_o})$.

The instance, $x2y2$, of composite function block type, $X2Y2$, in Figure 3, is represented as follows.

$$x2y2 = \langle x2y2, \phi, \phi, \{\langle DI1, 0 \rangle, \langle DI2, 0 \rangle\}, \{\langle DO, 0 \rangle\} \rangle \quad (6)$$

In above representation, the sets of output event and input data connections is represented as symbol $\phi$ (empty set), and those are determined when composite function block instance is connected with other.

2) Subapplication configuration algorithm

The IEC 61499 applications [18] are defined as network of function blocks that are connected with various types of function blocks, and since composite function block inner has network of function blocks that are connected with various types of function blocks, we can regard it as an application. In order to distinguish application that composes of inner function blocks of composite function block, from IEC 61499 applications, we will be called it subapplication.

Events and data of composite function block interface are connected with composite function block inner function blocks. Therefore, in order to define subapplication corresponding to composite function block, we must be defined the sets of type, instance, external input and external output events for service interface function blocks of start and stop that indicates start and end blocks of subapplication. Since only service interface function block type composes of function block type interface, interfaces of start and stop service interface function block types are defined as follows.

Definition 4(Interface of start and stop service interface function block)

Given composite function block interface, $g = \langle E_i, D_i, \omega_i, E_o, D_o, \omega_o \rangle$, respectively, interfaces of service interface function blocks for start and stop are a 6-tuple defined as

$$g^{START} = \langle \phi, \phi, \phi, E_i, D_i, \omega_i \rangle$$
$$g^{STOP} = \langle E_o, D_o, \omega_o, \phi, \phi, \phi \rangle.$$

Service interface function block instances for start and stop are defined such as composite function block interface.

Next, we add type and instance of service function blocks for start and stop to sets of types and instances of subapplication corresponding to composite function block. Therefore, set of type of subapplication composes of inner function block types, service interface function block types for start and stop, and set of instance composes of inner function block instances, service interface function block instances for start and stop.

Definition 5(Sets of external input and external output events)

Given type and instance of service function blocks for start and stop of subapplication, sets of external input and external output events of subapplication are defined as

$$I_e = \{\langle start, e \rangle | e \in E_o^{start}\},$$
$$O_e = \{\langle stop, e \rangle | e \in E_i^{stop}\}.$$

Where *start* and *stop* are instances of service function block types for start and stop.

A set of external input/output events is defined by service interface function block instances and the members in a set of external input events compose of a pair of instance name, its output event and values of related output data variables, and the members in a set of external output events compose of a pair of instance name, its input event and values of related input data variables.

Next, application state space that indicates execution state of subapplication corresponding to composite function block, is defined as follows.

Definition 6(Application state space)

When is given a set of function block types, $T(= T_b \cup T_c \cup T_s)$, and a set of function block instances, $I(= I_b \cup I_c \cup I_s)$, of subapplication, application state space, $S_a$, is defined as

$$S_a = 2^{S_s \cup S_b \cup S_c}$$
$$S_s = \bigcup_{k \in I_s} \left(\{n^k\} \times 2^{D_i^k} \times 2^{D_o^k}\right)$$
$$S_b = \bigcup_{h \in I_b} \left(\{n^k\} \times 2^{D_i^h} \times 2^{D_o^h} \times 2^{L^h} \times Q^h\right)$$
$$S_c = \bigcup_{j \in I_s} \left(\{n^j\} \times 2^{D_i^j} \times 2^{D_o^j}\right)$$

where radix $s, b, c$ denote service interface, basic and composite function block, respectively, and $n$ is function block instance name, $D_i$ is a value set of data input variables of function block, $D_o$ is a value set of data output variables of function block, $L$ is a value set of internal variables of basic function block, $Q$ is initial state of basic function block.

The Members of application state space represent all the states that all function block instances of application may have during execution. At any time, the state of basic function block instance is represented as a pair of instance name and current values of data input/output variables and internal variables, state name, and the state of service interface and composite function block instance is represented as a pair of instance name and current values of data input/output variables. Initial state in application state space composes of initial values of data input/output variables of each function block instances when is initialized to execute the application.

Initial state of application state space corresponding to subapplication of Figure 4 is represented as follows.

$$S_a^{X2Y2} = \{\langle x2y2\_start, \phi, \{\langle DI1, 0\rangle, \langle DI2, 0\rangle\}\rangle, \tag{7}$$

$$\langle add, \{\langle DI1, 0\rangle, \langle DI2, 0\rangle\}, \{\langle DO, 0\rangle\}, \phi, q_0\rangle, \tag{8}$$

$$\langle sub, \{\langle DI1, 0\rangle, \langle DI2, 0\rangle\}, \{\langle DO, 0\rangle\}, \phi, p_0\rangle, \tag{9}$$

$$\langle mul, \{\langle DI1, 0\rangle, \langle DI2, 0\rangle\}, \{\langle DO, 0\rangle\}, \phi, r_0\rangle, \tag{10}$$

$$\langle x2y2\_stop, \{\langle DO, 0\rangle\}, \phi\rangle\} \tag{11}$$

where $q_0, p_0, r_0$ are initial states of basic function blocks, $add, sub$ and $mul$.

The subapplication configuration algorithm corresponding to composite function block is as follows.

Input: composite function block type $cfbt = \langle g, T, I\rangle$

Output: subapplication, $a^c = \langle T^c, I^c, S_a^c, I_e^c, O_e^c\rangle$, corresponding to composite function block,

Process:

1: $g^{START} \leftarrow \langle \phi, \phi, \phi, E_i, D_i, \omega_i\rangle$ from $g$. // Get start SIFB type, $t^{START} = g^{START}$

2: $g^{STOP} \leftarrow \langle E_o, D_o, \omega_o, \phi, \phi, \phi\rangle$ from $g$. // Get stop SIFB type, $t^{STOP} = g^{STOP}$

3: Get the instances of start and stop SIFB types, $I^{start}$ and $I^{stop}$, respectively.

4: $T^c \leftarrow T \cup \{t^{START}\} \cup \{t^{STOP}\}$      // Get a set of types

5: $I^c \leftarrow I \cup \{I^{start}\} \cup \{I^{stop}\}$      // Get a set of instances

6: Get $I_e^c$ and $O_e^c$ for all SIFB instances of $T^c$.

7: Determine the initial states of application state space, $S_a^c$.

8: Returns the subapplication, $a^c = \langle T^c, I^c, S_a^c, I_e^c, O_e^c\rangle$.

3) Execution of composite function block

The event handling function of the composite function block instance that is used when composite function block is executed is defined next.

Definition 7(Event handling function of the composite function block instance)

Event handling function of the composite function block instance, $c^k$, is calculated by the following algorithm.

$$c^k : \langle s,t \rangle \rightarrow \langle s^+, t^+ \rangle$$

1: $\langle e,s,t \rangle \leftarrow e^k \langle s,t \rangle$     // Select the event of instance

2: $\langle I_e^k, s, t \rangle \leftarrow \langle e, s, t \rangle$     // Get a set of external inputs of composite function block instance

3: execute a subapplication, $a^k = \langle T^k, I^k, S_a^k, I_e^k, O_e^k \rangle$, corresponding to composite function block instance, $k$.

4: $for\ all\ e_o^k \in O_e^k\ do$

5:     $if\ e_o^k \neq \varepsilon\ then$     // If external output exists

6:         $\langle s,t \rangle \leftarrow o^k \langle e_o^k, s, t \rangle$     // event sending function of the instance

7:     $endif$

8: $endfor$

9: $\langle s^+, t^+ \rangle \leftarrow \langle s, t \rangle$

where $e^k$ is event selection function, $e_o^k$ is output event of composite function block instance, $k$, $o_e^k$ is a set of event outputs of composite function block instance, $k$, $o^k$ is event sending function.

Event handling function of the composite function block instance, when it received input event, gets external input values of composite function block instance from input event and executes subapplication corresponding to composite function block instance, triggers output events of composite function block instance corresponding to external output values of the subapplication.

Application step function that is called in application execution function [16] during execution of subapplication corresponding to composite function block is defined in detail next.

Definition 8(application step function)

Application step function, $p$, is calculated by the following algorithm.

$$p : \langle s,t \rangle \rightarrow \langle s^+, t^+ \rangle$$

1: $\langle k, s, t \rangle \leftarrow i \langle s, t \rangle$     // Select the instance

2: $if\ k \neq \varepsilon\ and\ type(k) \neq 'SIFB'\ then$

3:     $if\ type(k) \neq 'CFB'\ then$     // If instance is composite function block

4:         $\langle s,t \rangle \leftarrow c^k \langle s,t \rangle$     // Do event handling function of the composite function block

5:     $else$     // If instance is basic function block

```
6:       ⟨s,t⟩ ← h^k⟨s,t⟩         // Do event handling function of the basic function block
7:     endif
8:   endif
9: ⟨s⁺,t⁺⟩ ← ⟨s,t⟩
```

where $i$ is instance selection function, $type$ is function that returns type of function block instance, $c^k$ is event handling function of the composite function block, $h^k$ is event handling function of the basic function block.

Application step function is called at each repeat processing phase of application execution function and does event handling of the basic or composite function block instances, doesn't event handling of composite function block.

Next section shows computational experiment results for IEC 6149 applications with composite function block.

IV. Experiment results

This section shows first the results compared with previous models [16, 17] and next shows the experiment results tested in our distributed control program development tool, IIDesigner 1.0. Test was run on a Windows PC with in Intel Core i3 CPU 2.4GHz processor and 4GB RAM by method that measures and compares the execution time during 1 million cycles on the different applications. The motor operation example by the previous and proposed model using IIDesigner is shown in Figure 5 and Figure 6.

Table I shows comparison results on structural representation compared with previous model [18].

Table I. Comparison results on structural representation

| Method | BFB | SIFB | CFB | Application | Subapplication |
|---|---|---|---|---|---|
| Previous Model | ○ | ○ | × | ○ | × |
| Proposed model | ○ | ○ | ○ | ○ | ○ |

(○: supported, ×: not supported)

As shown in Table I, the proposed model can correctly represent the applications with hierarchical structure by presenting the formal definition on the composite function block.

Table II shows the functional comparison results compared with execution models, BSEM, NPMTR and CBEM, for IEC 61499 applications in [19]. As showed in Table II, the proposed model can correctly analyze the execution process for IEC 61499 applications with composite function blocks by defining newly the execution functions on the composite function blocks compared with previous models.

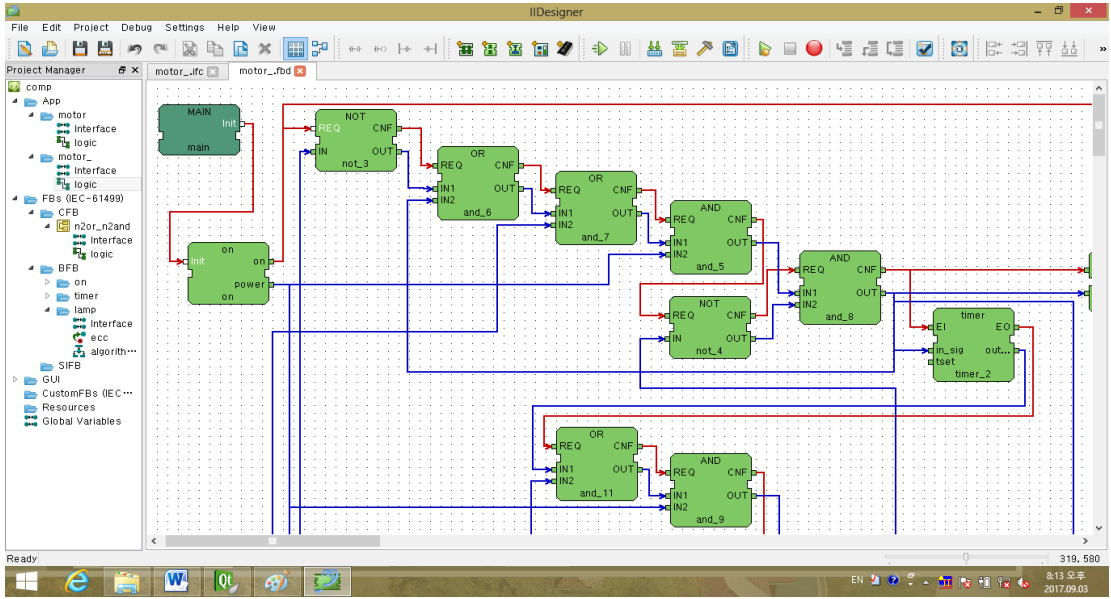

Figure 5. Motor operation example by previous model using IIDesigner

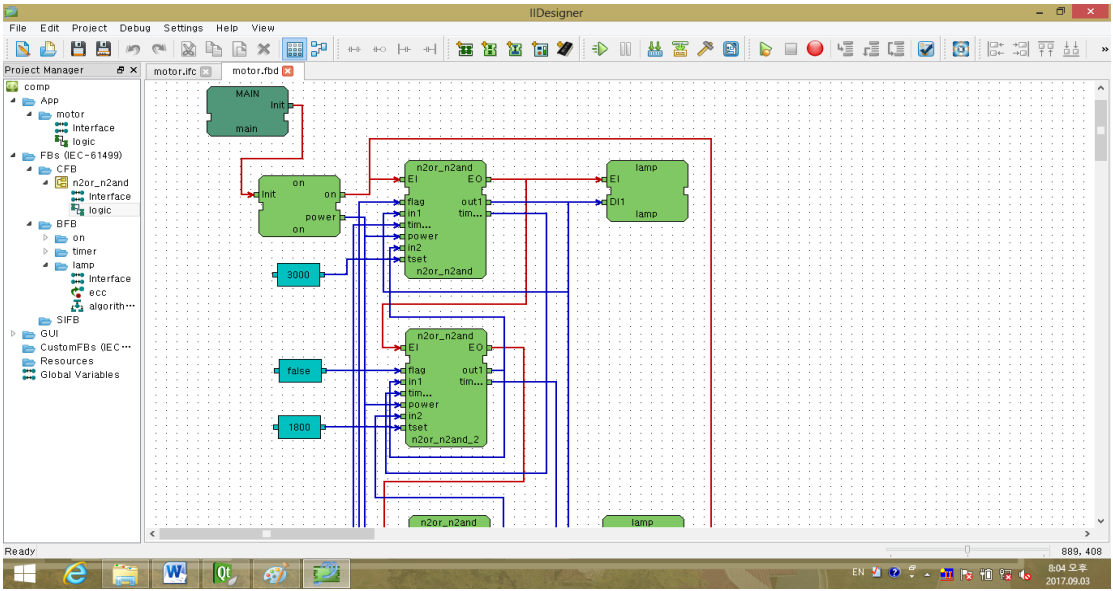

Figure 6. Motor operation example by proposed model using IIDesigner

Table II. Functional comparison results

| Method | Instance queue | Event queue | Boolean variable | BFB | SIFB | CFB |
|---|---|---|---|---|---|---|
| BSEM | ○ | ○ | × | ○ | ○ | × |
| NPMTR | × | × | ○ | ○ | ○ | × |
| CBEM | × | ○ | × | ○ | ○ | × |
| Proposed model | ○ | ○ | × | ○ | ○ | ○ |

Table III shows the size comparison results compared with previous models [18]. Columns 2-4, respectively, show the number of blocks and connections, and code size in previous models.

Columns 5-7, respectively, show the number of blocks and connections, and code size in proposed models.

Table III. Size comparison

| application | Previous model | | | Proposed model | | |
|---|---|---|---|---|---|---|
| | blocks | connections | Size (KBytes) | blocks | connections | Size (KBytes) |
| Motor operation | 26 | 46 | 3 | 9 | 18 | 2 |
| Fountain control | 43 | 79 | 4 | 21 | 42 | 3 |

As shown in Table III, the proposed model can simply represent the structure of application and reduce the size of code since number of function blocks and connections is small compared with previous models.

A comparison of execution time is shown in Table IV. As shown here, the execution time of the applications using proposed model is almost similar to previous model.

Table IV. Performance comparison (time in milliseconds)

| Method | Motor operation | Fountain control |
|---|---|---|
| Previous model | 54.0 | 206.4 |
| Proposed model | 55.0 | 213.2 |

Overall, the proposed model can be simply and hierarchically represented the distributed control applications and can reduce the size of execution program and can be effectively used in execution analysis of the applications composed of different kinds of function blocks including composite function block. A side-effect of using the proposed model is that a subapplication corresponding to composite function block must be configured. The IIDesigner compiler generates subapplication corresponding to composite function block, which is standard practice. Designers can change this subapplication by editing the composite function block. The use of subapplication corresponding to composite function block can increase runtime of software, but this is almost similar to previous model. However, it can reduce the compiled code size to use the composite function block and can be simply and hierarchically represented the software structure.

## V. Conclusion

Application of the distributed control system for factory automata composes of different kinds of function blocks. However, increment of the fields of application system complicates the structure and behavior of application extremely. Therefore, if it uses the proposed formal model for application, it extremely facilitates to perform the complex control logic and can improve the efficient of the software development. This paper has the advantage of as follows by presenting the formal definition on the composite function blocks.

First, the proposed model can simply represent the structure of the overall software of the distributed control systems.

Second, the proposed model can reduce the size of execution program.

Third, the proposed model can correctly analyze the execution process of the IEC 16499 applications with composite function blocks.

The proposed model will be effectively used to analyze runtime environment of the software tools for designing and developing of the complex application.